\documentclass{article}
\usepackage{graphicx}
\usepackage{amsmath}

\begin{document}

\title{Dewetting on porous media with aspiration}
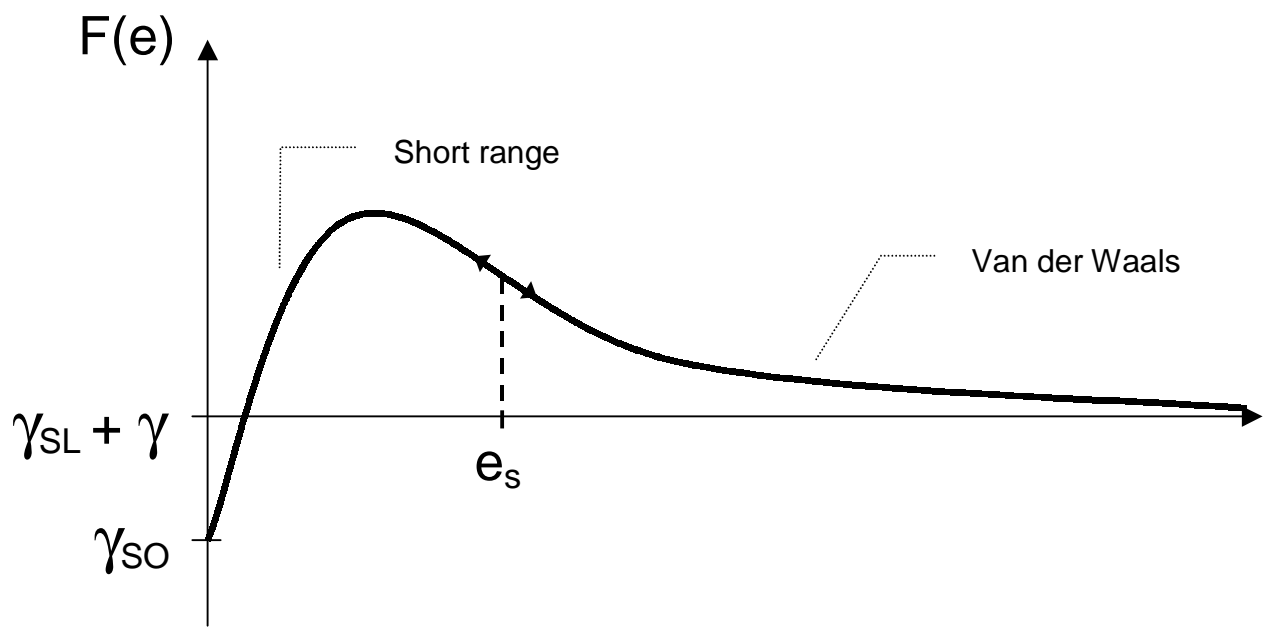
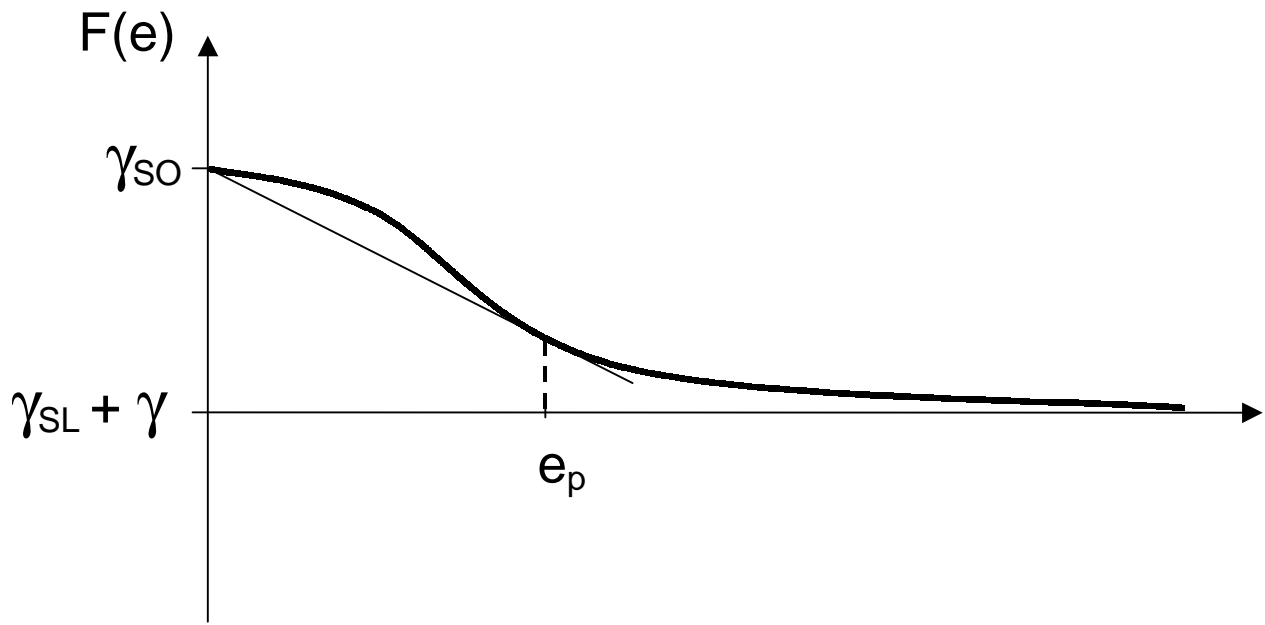
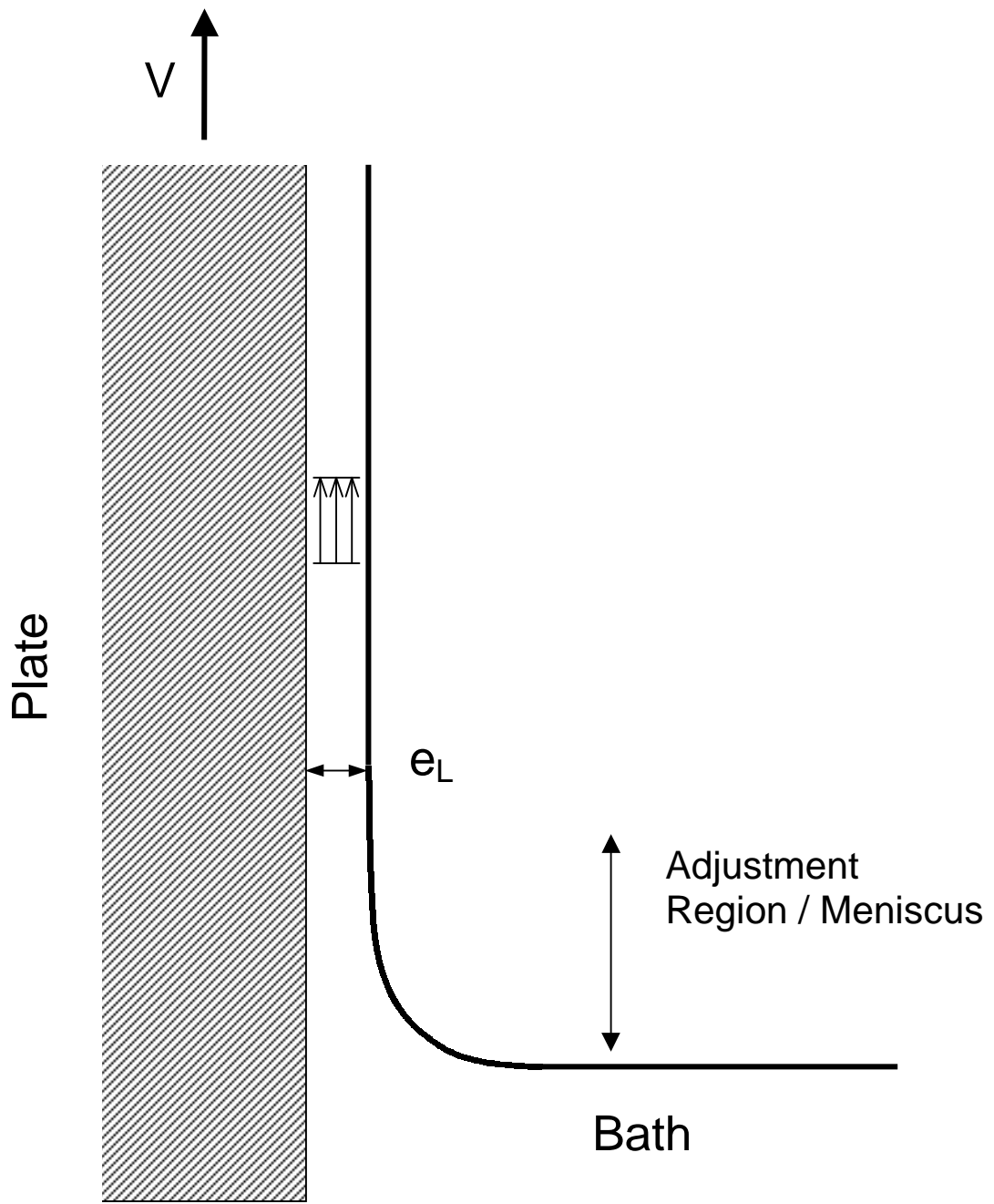
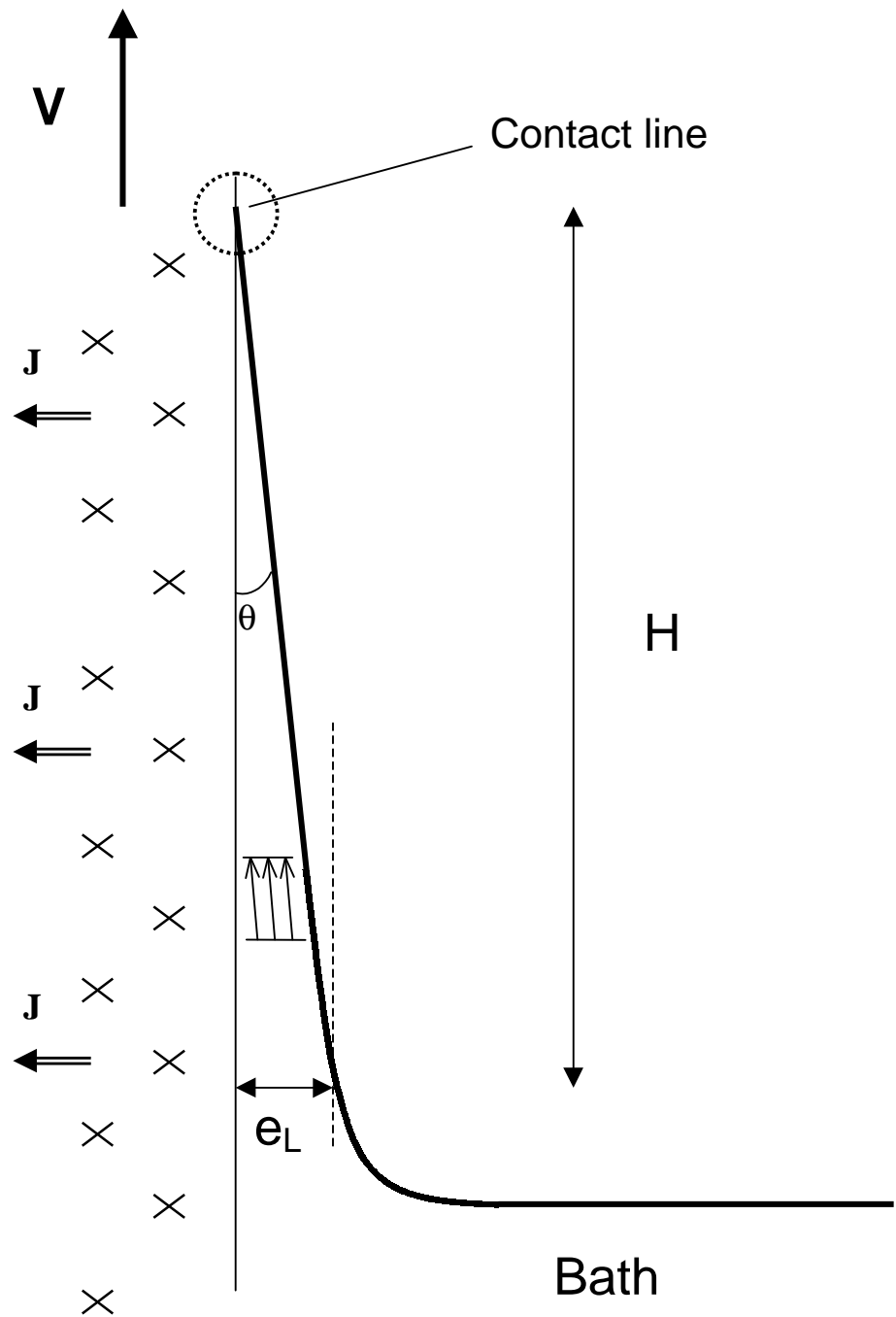
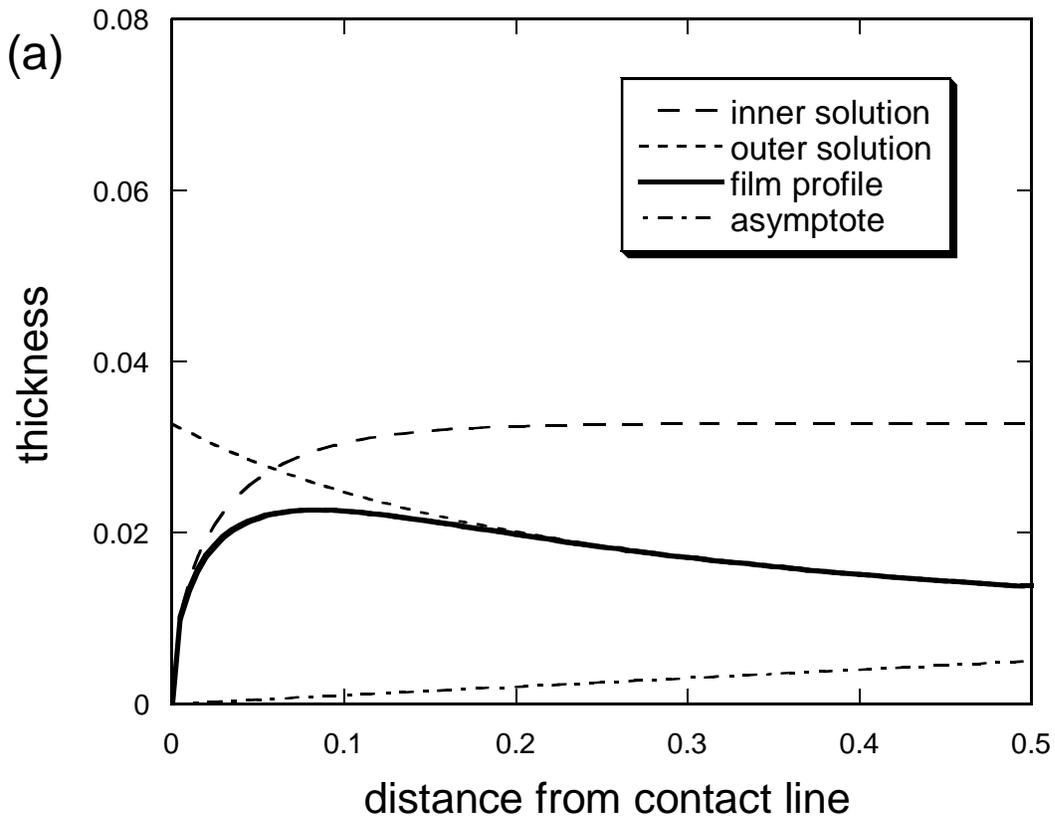
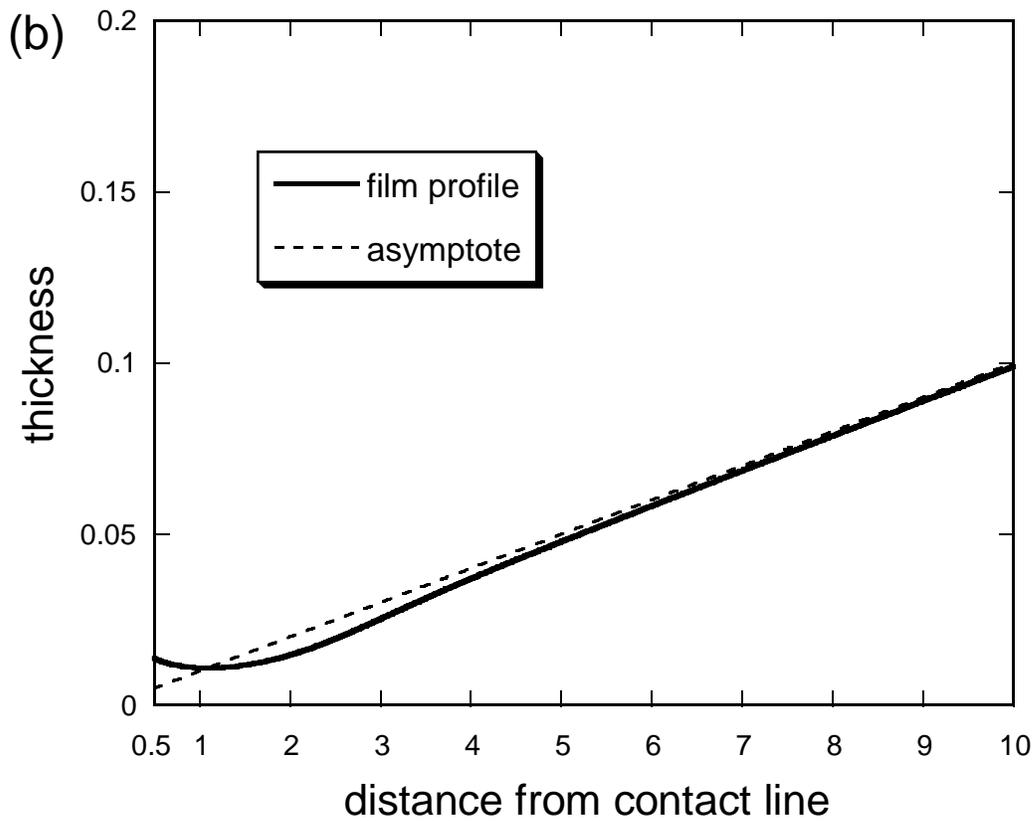
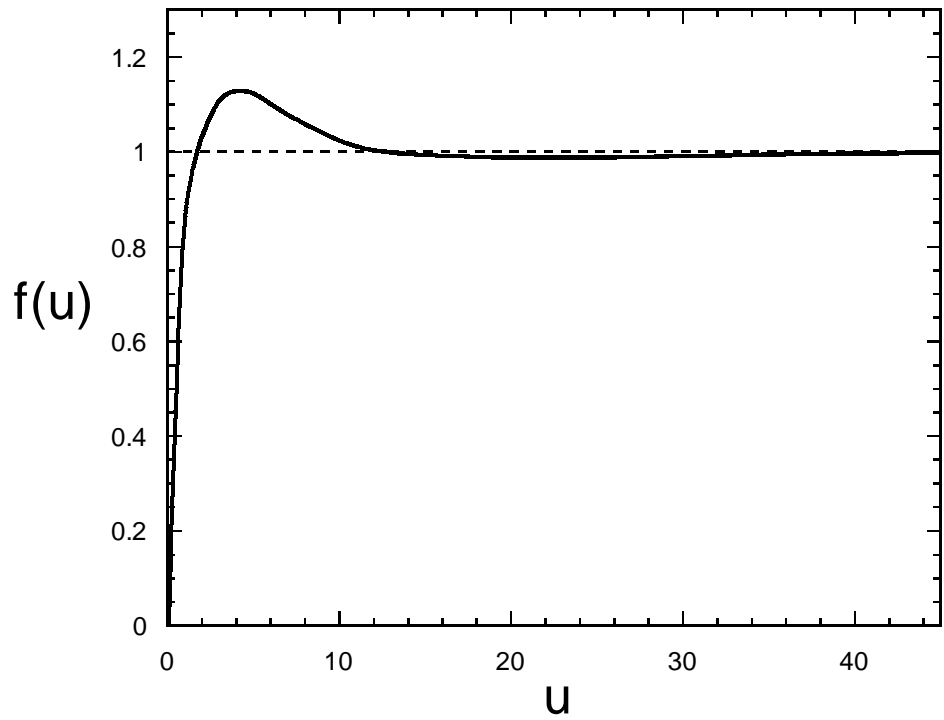
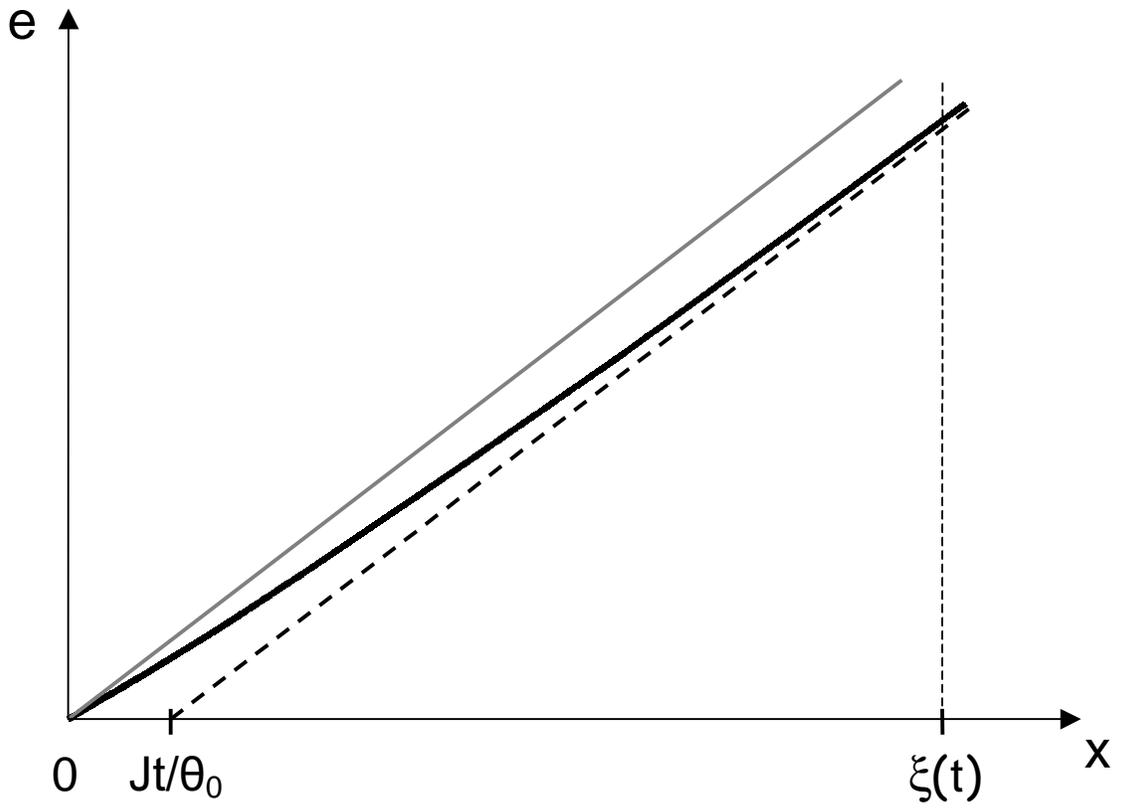
\author{A.\ Aradian, E.\ Rapha\"{e}l, P.\ G.\ de Gennes \\
Coll\`{e}ge de France, 11 place Marcelin Berthelot \and 75231 Paris Cedex
05, France \and E-mail: Achod.Aradian@College-de-France.fr,
 Elie@ext.jussieu.fr, \and Pierre-Gilles.deGennes@espci.fr}
\maketitle

\begin{abstract}
We consider a porous solid covered with a water film (or with a drop) in
situations where the liquid is pumped in, either spontaneously (if the
porous medium is hydrophilic) or mechanically (by an external pump).\ The
dynamics of dewetting is then strongly modified. We analyse a few major
examples \ \ a) horizontal films, which break at a certain critical
thickness \ \ b) the ``modified Landau-Levich problem'' where a porous plate
moves up from a bath and carries a film: aspiration towards the plate limits
the height $H$ reached by the film \ \ c) certain situation where the
hysteresis of contact angles is important.


\end{abstract}

\section{ General aims}

The current picture of dewetting processes on flat, smooth surfaces is
relatively clear~\cite{BrochardPGG92}, although inertial regimes still raise some
questions~\cite{Buguin}. But if the surface is porous, the situation is very
different. A number of practical problems belong to this class:

(i) in the 4-color offset printing, a sheet of paper may be at one moment
covered by a water film. The water must be removed very fast when the sheet
reaches the desired printing roller: this dewetting process is assisted by a
spontaneous suction of the water into the hydrophilic
paper~\cite{PGG94}.

(ii) many industrial processes involve the rapid passage of a solid film,
into a liquid bath.\ When the film leaves the bath, it drags a Landau-Levich
film~\cite{LandauLevich, Quere}. However, if the film is porous, and sucks in the fluid,
there is a height $H$ at which the Landau-Levich film disappears: what is
the value of $H$?~\cite{Elie}. \nopagebreak

(iii) when we paint with a brush, we are in fact transferring a liquid from
a porous medium onto a flat surface - the reverse\ process.

Our aim, in the present text is not to cover all these complex systems, but
to choose a few model examples where the effects of aspiration (towards the
solid) are non trivial.\ The essential parameter describing the aspiration
is the normal current $J$ (volume/unit area/unit time).\ When we pump through a
given thickness of porous solid, with a prescribed pressure drop, $J$ is
controlled by Darcy's law.\ When we have a spontaneous suction, over a time
interval $t$, $J$ is related to the Washburn equation~\cite{Washburn}: the scaling structure
of $J$ (for a pore volume fraction $\sim 1/2$ and wetting angles $\sim 1$
radian) is simply

\begin{equation}
\label{Washeq}
J \simeq \left( \frac{D}{t}\right) ^{1/2}.
\end{equation}

In eq.~(\ref{Washeq}), the parameter $D$ has the dimensions of a diffusion coefficient:

\begin{equation}
D=\frac{\gamma }{\eta }d\equiv V^{\ast }d,
\end{equation}
where $d$ is the pore diameter, $\gamma $ is the surface tension, $\eta $ the viscosity of the fluid,
and $V^{\ast }=\gamma /\eta$ is a characteristic capillary velocity. In all the model
systems to be discussed below, we shall assume that $J$ is prescribed and is
time independent. (One case where the time dependence may matter is
discussed in ref.~\cite{Elie}). We also assume that the pore diameter $d$ is very
small: then we may describe the capillary hydrodynamics in a continuous
picture.

One natural feature of a porous surface is to pin down a contact line - {\it i.e.} to
bring in a certain hysteresis in the contact angles. If the equilibrium
value of this angle (as given by Young's law) is $\theta _{e},$ the receding angle is $\theta
_{r}<\theta _{e}$. Two distinct cases can be found in practice:

\qquad a) `strong' pinning $\theta _{r}=0.$

\qquad b) `weak' pinning $\theta _{r}>0.$

In this paper, we start (sections~\ref{sec2} and \ref{sec3}) by a discussion of pumping effects
on surfaces with no hysteresis\footnote{ In order to have a good determination of the thermodynamic
 angle $\theta_{e}$, we do not necessarily need an ideal surface without defects. As shown by
 J.\ F.\ Joanny and P.\ G.\ de Gennes [J.\ Chem.\ Phys., \textbf{81}, 552 (1984)] in the case
 of ``regular'' defects, we need only a surface with irregularities below a certain strength
 threshold. In sections~\ref{sec2} and \ref{sec3},
 the local dissipation which might be associated to these defects is assumed
 to be negligible.}: this is not usual, but conceptually important. We then
proceed to some cases of hysteresis inducing weak or strong pinning.

In section~\ref{sec2}, we consider horizontal films, which thin out and become
unstable at a certain critical thickness. Then we proceed in section~\ref{sec3}
towards vertical plates pulled out of a liquid: the modified Landau-Levich
problem. In section~\ref{sec4}, we allow for hysteresis, first solving the macroscopic problem
of a pinned drop, then reconsidering the vertical plate situation. Section~\ref{sec5} discusses some
possible extensions and some limitations of the present work.

\section{Horizontal films}
\label{sec2}

\subsection{A macroscopic puddle under aspiration}

This case is described on fig.~\ref{puddle}. It has been studied recently
theoretically and experimentally by Bacri and Brochard~\cite{Bacri}. Here we
present only the case with no hysteresis, because it provides a good
introduction.

Let us denote by $S$ the spreading parameter $\gamma _{SO}-(\gamma +\gamma_{SL})$
where $\gamma $ is the liquid surface
tension, and $\gamma _{SL}$ (resp.\ $\gamma _{SO}$) the interfacial tension of the wet
(resp.\ dry) solid. We assume $S$ to be negative (partial wetting). The equilibrium
puddle has a thickness $e_{0}$ which results from a balance
between gravity (tending to decrease $e_{0}$) and capillarity (struggling
for a minimal exposed area). For a large
puddle (radius $R \gg e_{0}$) on a  partially wettable substrate, this
corresponds to the following energy:

\begin{equation}
\label{energy}
f=\frac{1}{2}\rho ge^{2}\frac{\Omega }{e} +\frac{\Omega }{e}\left( \gamma +\gamma
_{SL}-\gamma _{S0}\right)
\end{equation}
where $\Omega $ is the liquid volume, $\Omega /e$ the contact area, $\rho $
the density, and $g$ the gravitational acceleration. Minimizing~(\ref{energy})
with respect to $e$ gives a classical formula
for $e_{0}$ (which is typically of order 1mm)~\cite{PGGinRMP}:
\begin{equation}
e_{0}=\sqrt{2\frac{-S}{\rho g}}.
\end{equation}

Let us now pump on this structure, imposing a {\it small} current $J$ (and
assuming zero hysteresis).\ Then $e$ remains close to $e_{0}$. The horizontal
shrinkage velocity $U$ is related to $J$ by the conservation equation:

\begin{equation}
\label{shrinkage}
2\pi R e_{0}U=\pi R^{2}J,
\end{equation}
from which we deduce
\begin{equation}
U=J\frac{R}{2e_{0}}
\end{equation}

All the structure is slightly distorted. For instance the dynamic contact
angle $\theta _{d}$ is slightly smaller than $\theta _{e},$ because there is
a Poiseuille flow in the vicinity of the contact line.\ The necessity for
this flow can be perceived starting from the opposite assumption: if the
liquid in the puddle was flowing downwards, uniformly with velocity $J$, the
lateral velocity, deduced from the position of the contact line, would be
$\tilde{U}=J/\theta _{e}\neq U$. We can estimate the difference $%
\theta _{e}-\theta _{d}$ from a standard dissipation
argument~\cite{BrochardPGG92}:
\begin{equation}
T \overset{\bullet}{\Sigma} = \frac{3\eta }{\theta_{d} }(U-\tilde{U})^{2} \log \frac{e_{0}}{a} =%
(U-\tilde{U})\,\gamma \left( \cos \theta_{d} -\cos \theta _{e}\right)
\end{equation}
(where we have assumed $\theta _{e}$ and $\theta_{d} $ to be small, since this
is the most universal case). The argument in the logarithm is the size $\sim e_{0}
$ of the Poiseuille region (near the rim), divided by a molecular length.

\subsection{Nanoscopic pancakes or films}

We now pump on a very thin film ($e$ = nanometers), and assume first that the
solid surface is partly wettable ({\it i.e.} $\gamma _{SO}<\gamma _{SL}+\gamma $).
 At these small scales, the film energy $F(e)$ per unit area can be written as~\cite{Brochard91}
\begin{equation}
F(e)=\gamma _{SL}+\gamma +P(e),
\end{equation}
where, for $e$ much larger than the molecular size $a_{0}$, $P(e)$ is
controlled by long-range Van der Waals forces\footnote{ We only
retain non-retarded Van der Waals interactions: the retarded regime occurs for distances larger
than those that will be considered in this paper.}:
\begin{equation}
\label{VdWenergy}
P(e)=\frac{A}{12 \, \pi e^{2}} \hspace{1cm} (e \gg a_{0}).
\end{equation}
We will assume that the Hamaker constant $A$ is positive.
For shorter distances ($e \: \raisebox{-.4ex}{$\stackrel{>}{\sim}$} \: a_{0}$),
the precise form of $P(e)$ is generally unknown, but in the limit
$e \rightarrow 0$, we have $P(e \rightarrow 0)=\gamma _{SO}-\gamma -\gamma_{SL}$
since we must recover $F=\gamma _{SO}$ (dry solid).
The general aspect of $F(e)$ for partial wetting is shown on fig.~\ref{filmenergy}a.

Our film is locally
stable when the curvature $F''(e)$ is positive~\cite{Derjaguin}: thus, in the case of
fig.~\ref{filmenergy}a, we can get down without changes to a certain thickness
$e_{s}$, such that $F''(e_{s})=0$.
But, when $e$ reaches $e_{s}$, a spinodal decomposition is expected. If we
stopped the pumping at this moment, this would ultimately lead to a
coexistence between dry regions and finite, macroscopic droplets. But if we persist in
pumping, the early droplets (of thickness $\sim e_{s}$) will fade out before
any major coalescence process.

A similar discussion can be given for the case of complete wetting ($\gamma
_{SO}>\gamma _{SL}+\gamma $). The corresponding plot of $F(e)$ is described
in fig.~\ref{filmenergy}b. In the absence of pumping, the liquid forms a ``pancake''
of {\it nanoscopic} thickness $e=e_{p}$.
The equilibrium condition determining $e_{p}$ is~\cite{PGGinRMP}
\begin{equation}
\label{equilibrium}
e_{p}\Pi (e_{p})+P(e_{p})=S
\end{equation}
where $\Pi (e)\equiv -\partial P/\partial e$. Eq.~(\ref{equilibrium}) may be interpreted
as a balance of forces on the contact line: the
Young force $S=\gamma _{SO}-(\gamma _{SL}+\gamma )$ is equilibrated
by the pressure due to Van der Waals attractions. (The graphical
construction of $e_{p}$ is shown on fig.~\ref{filmenergy}b: $e_{p}$
is the point for which the tangent to $F(e)$ intercepts the vertical
axis at $\gamma_{SO}$.)

As a consequence of this, when pumped, a continuous film would tend, at $e<e_{p}$,
to break into a mixture of such pancakes and dry regions rather than thin out uniformly.
Each pancake will then shrink by a process reminiscent of the puddle
[{\it e.g.} eq.~(\ref{shrinkage})].

Of course, all this discussion of nanoscopic objects is rather unrealistic,
since most porous media have pores larger than the structure described here:
our continuum description can be adequate only if our porous surface was
based on an extreme permeation membrane, with pore diameters of a few
angstr\o ms! But, in spite of this strong limitation, the discussion is (we
think) conceptually helpful.

\section{The modified Landau-Levich problem}
\label{sec3}

\subsection{A reminder on non porous systems}
\label{reminder}

We pull out a vertical plate from a bath (fig.~\ref{plate}a) at velocity $V$. If the
plate is partially wettable (equilibrium contact angle $\theta _{e}),$ and
the velocity $V$ is very low, we retain a dynamic contact angle $\theta _{d}$
close to $\theta _{e}.$ If we increase the velocity $V$, we have a non trivial
relation $V(\theta _{d})$. Of course $V(\theta _{d}$) vanishes at $\theta
_{d}=\theta _{e}$, but $V$ also vanishes at $\theta _{d}\rightarrow 0$,
because the dissipation in a thin wedge is very large. The simplest
(crudest) form, valid for small angles, is~\cite{BrochardPGG92}
\begin{equation}
V(\theta _{d})=const. \; V^{\ast }\left( \theta _{e}^{2}-\theta
_{d}^{2}\right) \theta _{d}.
\end{equation}
This has a maximum:
\begin{equation}
V=V_{m}=const. \; V^{\ast }\theta _{e}^{3}.
\end{equation}

If we impose a velocity larger than $V_{m}$ (as usual in fast technical
processes) we cannot retain a contact line: the plate drags with it a film
of finite thickness $e_{L}$.\ This thickness was computed in a classic paper
by Landau and Levich~\cite{LandauLevich} (for viscous regimes):
\begin{equation}
\label{thicknessLL}
e_{L}=b\left( \frac{V}{V^{\ast }}\right) ^{2/3}
\end{equation}
where $b$ is related to the curvature of the underlying meniscus ($b\sim 1$ mm).
The height of the adjustment region (beyond which the film becomes
uniform) is of order $b \: (V/V^{\ast})^{1/3}$.

The above results for the film thickness
 and the adjustment region are also valid for a completely wettable plate. In that case,
  however, there is no velocity threshold and the plate drags a film as soon as $V>0$.

Eq.~(\ref{thicknessLL}) has been amply verified, and extended up to inertial regimes~\cite{Quere}.
Our aim here is to see how the film is modified under pumping, in
a viscous regime of small capillary numbers ($C\!a=V/V^{\ast} \ll 1$).

\subsection{Pull out of a surface under pumping: Macroscopic flow}
\label{macroscopicsection}

We now consider the pull-out of a substrate able to suck liquid in as
it drags a film from the bath. This {\it modified} Landau-Levich problem
was discussed in a simplified form in ref.~\cite{Elie}, and, for completeness,
we start by a brief summary of the main results.

Let us assume that we impose a uniform pumping current $J$ on the film,
$J$ being small compared to the drag velocity $V$. The aspect of the flow
is pictured on fig.~\ref{plate}b: as in the usual Landau-Levich
film, there is a meniscus from the bath and an adjustment region
leading to a thickness $e_{L}$. But then on, aspiration comes
into play, and we expect that most of the film will have a
simple hydrodynamic flow (plug flow) in the external film, without
any pressure gradients, where the flow lines are just tilted by the
angle $\theta=J/V \: \ll 1$. As a consequence, the Landau-Levich film thins out and ultimately
disappears at a finite height $H=e_{L}/\theta$.

However the validity of this non-dissipative solution is necessarily restricted to macroscopic
scales, and is not acceptable very close to the
contact line, where one must take into account local wetting properties such as the spreading
parameter $S$: the slope of the profile must change.
It is our aim in the rest of this section to understand the
structure of the film in the vicinity of the contact line, both in the case of complete
($S \geq 0$) and partial wetting ($S < 0$).

\subsection{Hydrodynamic equation and boundary conditions}
\label{hydroeq}

Let us denote $z$ the vertical distance measured downwards from the contact line.
 In the lubrication approximation, we can describe the velocity in the film
 (in the $z$-direction) by the following parabolic form:
\begin{equation}
\label{Poiseuille}
v_{z}(x)=-V+\frac{V_{1}}{e^{2}}(x^{2}-2ex),
\end{equation}
dictated by the no-slip condition on the plate, and the absence
of tangential stress at the free surface (note that in this formula,
 $V_{1}$ and $e$ are functions of $z$).

For a steady state situation, conservation imposes that the upwards flow rate $-\int_{0}^{e}v_{z}(x)dx$
 at height $z$ balances the amount $Jz$ of liquid that disappears above due to the pumping:
\begin{equation}
\label{conservation}
Jz=-\int_{0}^{e}v_{z}(x)dx=e \left(V+\frac{2}{3}V
_{1}\right).
\end{equation}
The Poiseuille flow~(\ref{Poiseuille}) is driven by a pressure gradient $\partial
p/\partial z$, where $p$ contains contributions from capillarity and Van der
Waals forces
\begin{equation}
p=-\gamma \frac{\partial ^{2}e}{\partial z^{2}}-\frac{A}{6\pi
e^{3}},
\end{equation}
and $A$ is a Hamaker constant.

Writing
\begin{equation}
\frac{\partial p}{\partial z}=\eta \frac{\partial^{2} v_{z}}{\partial
x^{2}}=\frac{2\eta V_{1}}{e^{2}},
\end{equation}
equation~(\ref{conservation}) can be rewritten as

\begin{equation}
\label{centraleq}
1-\frac{\theta z}{e}=\frac{e^{2}}{3\eta V}\left\{ \gamma \frac{d^{3}e%
}{dz^{3}}+\frac{d}{dz}\left( \frac{A}{6\pi e^{3}}\right) \right\}
\end{equation}
where, as before, $\theta =J/V$ is a small angle ($\theta \ll 1)$.

Equation~(\ref{centraleq}) is reminiscent of the equation discussed in ref.~\cite{PGGinRMP} for precursor
films over a wettable surface, and in ref.~\cite{XiaPGG} for an advancing liquid on a
 partly wettable surface. But there are two differences: a)
pumping introduces a new term ($-\theta z/e$) on the left hand side, b)
the direction of the driving velocity $V$ is reversed (in the frame
of the plate, the liquid front moves back).

In order to single out the physical solution of eq.~(\ref{centraleq}), it is necessary
to impose the three following boundary conditions:

\begin{eqnarray}
\label{condition1}
i)&& e(z)=0 \mbox{\qquad at } z=0  \ ; \\
\label{condition2}
ii)&& \lim_{z \to 0} \; \frac{A}{12 \pi \: e(z)^2} - \frac{1}{2} e'(z) = \frac{S}{\gamma}
\ ; \\
\label{condition3}
iii)&& e \mbox{ reaches a linear profile when $z$ goes to infinity}.
\end{eqnarray}
The boundary condition ($ii$) reflects the local balance of forces
on the contact line~\cite{PGGinRMP}. Since it involves the spreading
parameter $S$, it will introduce a distinction between situations of
total wetting ($S>0$) and situations of partial wetting ($S<0)$.

Far from the contact line, as expected, the profile merges into the
macroscopic profile of angle $\theta$ described in the previous section:
in eq.~(\ref{centraleq}), for large thicknesses and vanishing curvature,
$e \sim \theta z + C$ is easily seen to be
the asymptotic profile ($C$ a constant). But a more thorough
analysis of the asymptotic behaviour of eq.~(\ref{centraleq})~\cite{technical} shows
that any profile of angle $\theta$ is {\it not} convenient, and that
we must necessarily set $C=0$. Hence, remarkably enough, if the asymptotic profile
were known with sufficient accuracy, by a straight extrapolation of it one could guess
the actual microscopic position of the contact line, regardless of any complicated local
behaviour.

Bearing this essential fact in mind, we now turn to solving eq.~(\ref{centraleq})
in the case of a completely wettable substrate.

\subsection{Wettable surface}
\label{wetting}

We first consider the modified Landau-Levich problem on a wettable
surface. From now on, to make later explanations clearer, the two terms
of the l.h.s.\ of eq.~(\ref{centraleq}) will be respectively referred to as the
`dynamic term' (since the presence of the ``$1$'' is directly due to the pull-out)
and the `aspiration term', and the terms on the r.h.s.\ respectively as `capillary'
and `VW' (Van der Waals) terms.

We start from the `far-field' region and look for characteristic
lengths to scale eq.~(\ref{centraleq}). In these scales, we need both the dynamic
and aspiration terms to be of order one (to give rise to the asymptotic profile of angle
$\theta$), plus a third term to describe the departure from it. One can show that the only
consistent choice is to keep the VW contribution of order one. This
imposes the following scalings:
\begin{eqnarray}
\label{defz1}
z_1  = z / \mu_1 & \mbox{with } & \mu_1 = \ell \\
\label{defe1}
e_1  = e / \lambda_1 & \mbox{with } & \lambda_1 = \theta \ell \,
\end{eqnarray}
where $\ell$
is a crucial length of the problem defined by
\begin{equation}
\label{ell}
\ell = a \left( \frac{V^{\ast }}{J} \right)^{1/2}
\end{equation}
and $a$ an atomic length (of a few \AA) defined as $a^{2}=\frac{A}{6\pi \gamma}$.
We note that $\ell$, though much greater than $a$, remains in the microscopic domain
($J/V^{\ast}=10^{-4}$ gives a few hundred \AA).

Our central equation~(\ref{centraleq}) rewrites in these
scaled variables
\begin{equation}
\label{outereq}
1-\frac{z_1}{e_1} = \epsilon \: e_1^2e_1''' - \frac{e_1'}{e_1^2} \ ,
\end{equation}
where
\begin{equation}
\epsilon = \frac{1}{3} \: \frac{V^{\ast}\theta^3}{V}
\end{equation}
is a small parameter of the problem\footnote{ To put numbers, taking $\theta=10^{-2}$,
$V=0.7\ $m.s$^{-1}$ and $V^{\ast}=70\ $m.s$^{-1}$, gives $\epsilon \simeq 3.10^{-5}$.}:
$\epsilon \ll 1$. We thus see that in these scales, capillarity appears as a
perturbative term to a behaviour ruled mainly by the pull-out velocity,
aspiration and VW. However, we immediately notice that the
capillary term, though of order $\epsilon$, contains the highest derivative in the
equation, implying that we are in presence of a \emph{singular perturbation} problem.

Such problems have the interesting property
that even for arbitrary small $\epsilon$, there always is a
region, called \emph{inner} region or \emph{boundary layer}, where
the true solution dramatically differs from the simple $\epsilon=0$ solution\footnote{
Or even from any perturbative expansion in the powers of $\epsilon$.}
(usually named \emph{outer} solution). For the interested reader, extensive coverage of
this subject can be found in refs.~\cite{Bender,VDyke,Nayfeh,Kevorkian}.
In the following presentation, we will try to emphasize physical content rather than enter
into technical developpements~\cite{technical}.

In our problem, the inner region stands near the contact line, at
$z=0$, and capillarity is called to play an important role there. The outer region,
on the other hand, extends from the limit of the inner region (crudely) to infinity.

The (first-order) outer solution is looked for by setting $\epsilon=0$
in eq.~(\ref{outereq}). We obtain the exact form
\begin{equation}
\label{outersolution}
e_{\, 1,\mathrm{\, out}}=\frac{e^{z_1^2/2}}{K + \int_{0}^{z_1}{e^{y^2/2}\, dy}}
\: ,
\end{equation}
where K is a constant that will be determined later by
matching with the inner solution. One can check that this outer solution
respects the asymptotic boundary condition~(\ref{condition3}).

In the inner region, the essence of singular perturbation techniques
is to rescale the independent variable with the help of a power of
$\epsilon$ in order to change the dominant balance of the equation. Guided by the form
of the boundary condition~(\ref{condition2}) which balances capillarity and VW at the contact
line, we choose the following new variables:
\begin{eqnarray}
z_2  = z / \mu_2 & \mbox{with } & \mu_2 = \epsilon^{1/2} \: \ell \\
e_2  = e_1 \ \ \ & & \lambda_2=\lambda_1 \ .
\end{eqnarray}
The expression of $\mu_2$ reveals that the boundary-layer near the
contact line extends over a distance of order $\epsilon^{1/2} \:
\ell$.

In these inner variables, eq.~(\ref{centraleq}) takes the inner
form
\begin{equation}
\label{innereq}
\epsilon^{1/2} - \epsilon \, \frac{z_2}{e_2} = e_2^2e_2''' -
\frac{e_2'}{e_2^2} \ ,
\end{equation}
where the dynamic term and the aspiration term remain small, and the behaviour is dictated
mainly by a balance between capillarity and VW. Solving to first order (by setting $\epsilon = 0$,
and integrating twice), we get $z_2$ as a function of $e_2$:
\begin{equation}
\label{innersolution}
z_{\, 2,\mathrm{\, in}}= \int_{0}^{e_2} {\frac{\tilde{e} \, d\tilde{e}}{\sqrt{ C\,\tilde{e}^3
+ 2 D\,\tilde{e}^2 + \frac{1}{3}}}} \ .
\end{equation}
The lower limit of integration in~(\ref{innersolution}) is set to $0$ by the boundary
condition~(\ref{condition1}), while application of~(\ref{condition2})
brings
\begin{equation}
\label{D}
D=-\frac{1}{3} \: \frac{V^{\ast}J}{V^2} \: \frac{S}{\gamma}
\end{equation}
(negative for our wettable surface).

We now proceed to the asymptotic matching~\cite{Bender,VDyke,Nayfeh,Kevorkian} of the inner and the
outer solution. The first order matching condition writes
\begin{equation}
\label{matchingcondition}
\lim_{z_2 \to +\infty} e_2(z_2)=\lim_{z_1 \to 0} e_1(z_1).
\end{equation}
The l.h.s.\ admits a finite limit only if $C$ takes the specific value
\begin{equation}
\label{C}
C = \frac{2 \sqrt{2}}{3}(-D)^{3/2}.
\end{equation}
The value of this limit is then $e_p/\theta \ell$, where $e_p$ is the
thickness of the static pancake [eq.~(\ref{equilibrium})] that the liquid would
form at rest on the surface, without aspiration. We then equate this
value with the r.h.s.\ limit of the matching condition, which is equal to $1/K$,
and find:
\begin{equation}
\label{K}
K=\frac{\theta \ell}{e_{p}}.
\end{equation}

The solutions found for the inner [eq.~(\ref{innersolution})] and outer
[eq.~(\ref{outersolution})] regions, together with eqs.~(\ref{D}),~(\ref{C})
and~(\ref{K}), give us a complete description
of the profile of the film over a wettable substrate. Figures~\ref{wettable}a and~\ref{wettable}b
present the profile that is obtained, respectively in the close vicinity of the contact
line, and farther away in the outer region.

Our results call for a few important comments:

i) When plotted in orthonormal coordinates, the liquid
takes the shape of a long, almost flat, strip at the approach of the
contact line. This is naturally a remembrance of the usual precursor
film that is encountered in the wetting process of a (wettable) surface by
a drop~\cite{PGGinRMP}. In physical units, the thickness of this
tongue is indeed similar to that of a precursor, both being of order
$e_p$.\footnote{ The appearance of a static quantity as $e_p$ is not a
surprise: in the inner region, the dynamic term was totally
ignored. If we increase the dynamic drive (by
increasing $V$), this static approximation can still be made, but
the region over which it remains valid ($\sim \epsilon^{1/2}
\ell$) diminishes accordingly.}

ii) We are now able to give an order of magnitude for the distance
over which the flow merges with the simple, non-dissipative, flow
of sec.~\ref{macroscopicsection}. This merging is described by the
outer solution $e_{\,1 , \, \mathrm{out}}$, and in fig.~\ref{wettable}b we see that it
occurs over a distance of a few $\ell$'s~[eq.(\ref{ell})], that is to say on
microscopic scales. Practically, only the macroscopic tilted flow of angle $\theta$
is observable, thus validating the simple analysis of
ref.~\cite{Elie}.\footnote{ In the limit $J \to 0$, the merging length ($\sim \ell$)
increases towards infinity, but one must notice that at the same time $\theta \to 0$ at a
faster rate: there is no contact line anymore, and we go back to a classical infinite
Landau-Levich film.}

iii) One question remains: what determines the actual position of
the contact line ? We know from the previous section that a
delicate hydrodynamic balance entails that the contact line
necessarily lies on the position obtained by a straight extrapolation
of the asymptotic linear profile of angle $\theta$. If this
asymptotic line is shifted, then the whole profile is shifted by
the same amount, without change: all the positions of the contact
line are equivalent. But this degeneracy is naturally broken when
we attempt to match the asymptotic profile with the meniscus originating
in the bath from which the plate is pulled out. This matching,
by fixing the asymptotic line, also determines the contact line position.

\subsection{ Partially wettable surface}
\label{partialwetting}

We now consider the situation of a  partially wettable surface. In fact, physical
intuition encourages us to think that this will not be a
mere extension of the wetting case. In the pull-out process, due
to aspiration, the contact line is forced to recede on the plate.
For a wettable substrate, we have shown that this dewetting occurs
with the liquid leaving, just after the contact line, a thin strip similar
to a precursor film. On a  partly wettable surface, however, the liquid is not
inclined to leave any film at all, but rather to adopt a sharp
profile which is in obvious contradiction with the asymptotic line
of small angle $\theta \ll 1$. How does the system resolve the
dilemma ?

We analyse this situation along the same lines as previously.
In the $(z_1,e_1)$ set of variables [eqs.~(\ref{defz1}) and (\ref{defe1})], capillarity is still
a singular perturbative term, but is called to be important near the contact line.
The solution will consequently still exhibit a boundary-layer
structure: the outer solution [eq.~(\ref{outersolution})] must be
abandoned near in the vicinity of the contact line, where the inner
solution given by eq.~(\ref{innersolution}) takes place.

However, in eq.~(\ref{innersolution}), an important change occurs:
the constant $D=-\frac{1}{3} \: \frac{V^{\ast}J}{V^2} \:
\frac{S}{\gamma}$ [eq.~(\ref{D})] is now \emph{positive}, because
the spreading parameter $S$ is negative in the  partial wetting
case. We then go on and try to match the inner and the outer
solution with the condition~(\ref{matchingcondition}), but this
appears to be impossible: whatever the value of the constant $C$ in
the inner solution, the l.h.s.\ of eq.~(\ref{matchingcondition})
does not admit any finite limit.\footnote{ It can be shown that
even with the help of more refined matching procedures than that
given by eq.~(\ref{matchingcondition}), no matching can be performed between
the inner and the outer solution.}

This mismatch takes its roots in the discrepancy between the
behaviour of the profile in the inner region, where it
is similar to the steep, hyperbolic profile of a static drop on a
partially wettable surface~\cite{PGGinRMP}, and the outer region where the
profile tends to join the low-angle asymptote. Therefore, a third region, called
\emph{intermediate region}, must appear between the inner region and the outer region,
so as to bridge the gap.

For the intermediate region to play its role, one is led to keep
capillarity, VW and the dynamic term of order one. To achieve
this, both the independent and the dependent variables must be
scaled with appropriate powers of $\epsilon$. We find:
\begin{eqnarray}
\label{defz3}
z_3  = z / \mu_3 & \mbox{with } & \mu_3 = 3^{-1/6} \, \epsilon^{1/6} \, \mu_1 \\
\label{defe3}
e_3  = e / \lambda_3 & \mbox{with } & \lambda_3 =3^{1/6} \, \epsilon^{-1/6} \, \lambda_1,
\end{eqnarray}
so that the `intermediate equation' takes the form
\begin{equation}
\label{intermediateeq}
1 - \epsilon^{1/3} \, \frac{z_3}{e_3} = e_3^2e_3''' -
\frac{e_3'}{e_3^2}\ .
\end{equation}

Deriving an analytical closed-form solution of this strongly non-linear
equation reveals to be a difficult task. However, a satisfying
approximate solution can be constructed from a linearized
form of eq.~(\ref{intermediateeq}), and be matched through
a numerical procedure with both the outer and inner solutions to determine all the
constants that remained undetermined (yielding values of $C$ and $K$ that are close to $0$).
\footnote{ For the sake of conciseness, a technical presentation of the
procedure~\cite{technical}, unessential to understand the physical picture, has been
omitted here.}
As can be seen on fig.~\ref{partialwetting}, the profile now presents a `bump' that
allows a transition between the two extremal regions.

Let us focus our attention on the evolution of this bump when $\epsilon$
becomes smaller, for instance when increasing the pull-out velocity
$V$. The length and thickness of the bump are
respectively of order $\mu_3 = 3^{-1/6} \, \epsilon^{1/6} \, \mu_1$
and $\lambda_3 =3^{1/6} \, \epsilon^{-1/6} \, \lambda_1$. Therefore, compared to the
characteristic lengths $\mu_1$ and $\lambda_1$ in the outer region,
the bump's length shrinks, while, as is shown in fig.~\ref{bumpsize}, it gets steeper  because
of the negative power of $\epsilon$ appearing in $\lambda_3$. Such a steepening
appears coherent as an attempt to reconcile conflicting evolutions:
when $V$ is incremented, the asymptotic profile flattens more and more ($\theta \to 0$),
in increasing contradiction with the `natural' tendency of the profile to display strong
slopes in the inner region.

We conclude by pointing out that, as in the wetting case, the merging distance
with the macroscopic flow of angle $\theta$ is of order $\ell$: for practical
purposes, there is no visible deviation from the uniformly tilted flow.
Indeed, it is an interesting fact that, except for a small region
near the contact line where the profile is submitted to the influence of wetting
properties (as the spreading parameter), most of the flow is
exactly the same in the wetting and the partial wetting case. This is
of course to be imputed to the far off-equilibrium issue that
we are dealing with: aspiration by the plate forces a contact
line to appear in situations where the pull-out is so strong that
there should not exist any.

\section{Effects of hysteresis}
\label{sec4}

\subsection{Pinned, macroscopic, horizontal drop}

Let us consider a macroscopic drop on top of the horizontal upper
surface of a porous medium with hysteresis.
The drop is submitted to a uniform
downwards aspiration (current $J$) while its contact line is strongly pinned
($\theta_{r}=0$) at $x=0$ (the $x$-axis is along the horizontal surface).
Initially, the (macroscopic) profile near the contact line, $e$, is
wedge-like (with a contact angle $\theta_{0})$: $e(x,t=0)=\theta_{0}x$.
How does the profile evolve under the combined effect of
aspiration (which would force the contact line to recede) and
pinning (which prevents it from doing so) ?

In the lubrication approximation, the horizontal velocity field inside the drop
writes, using a vertical $z$-axis:
\begin{equation}
v(x,z)=\frac{U}{e^{2}}\left( z^{2}-2ez\right)
\end{equation}

This Poiseuille flow is driven by a pressure gradient $\partial p/\partial x$
where $p=-\gamma \partial^{2}e/\partial x^{2}$ (for the macroscopic
drop under consideration, Van der Waals forces can be ignored,
except in a very small region near the contact line).

Writing $\partial p/\partial x=\eta \partial ^{2} v/\partial
z^{2}=2\eta U/e^{2}$, we arrive at
\begin{equation}
-\gamma \frac{\partial^{3}e}{\partial x^{3}}=2\eta
\frac{U}{e^{2}}.
\end{equation}
The horizontal flow rate, $Q=\int_{0}^{e} v(z)\,dz$, is therefore given by
\begin{equation}
\label{flowrate}
Q=-\frac{2}{3}U e=\frac{1}{3}V^{\ast} e^{3}
\frac{\partial^{3}e}{\partial x^{3}}.
\end{equation}

Inserting eq.~(\ref{flowrate}) into the conservation equation
\begin{equation}
\frac{\partial e}{\partial t} + J + \frac{\partial Q}{\partial
x}=0,
\end{equation}
we find the evolution equation of the profile:
\begin{equation}
\label{evolution}
\frac{\partial e}{\partial t} + J +\frac{\partial}{\partial
x}\left(\frac{1}{3} V^{\ast} e^{3}
\frac{\partial^{3}e}{\partial x^{3}}\right)=0.
\end{equation}

For a small pumping current J, $e$ remains close to $\theta_{0}x$
and eq.~(\ref{evolution}) can be linearized:

\begin{equation}
\label{evolutionlin}
\frac{\partial e}{\partial t} + J +\frac{\partial}{\partial
x}\left(\frac{1}{3} V^{\ast} \theta_{0}^{3}x^{3}
\frac{\partial^{3}e}{\partial x^{3}}\right)=0
\end{equation}
(one can check that the above simplification is actually valid for $J \ll V^{\ast}
\theta_{0}^{4}$).

Far away from the contact line, we expect that the profile is not
significantly different from that obtained in the
absence of pinning, which would merely be a global downwards translation
of the initial wedge by a distance $Jt$ (yielding $e = \theta_{0}x - J t$).
It is therefore natural to
look for a solution to eq.~(\ref{evolutionlin}) of the form
\begin{equation}
e(x,t)=\theta_{0} x -J t \; F(x,t),
\end{equation}
where the function $F$ describes the effect of the pinning, and is such that,
 at any given time $t$, $F(x,t) \sim 1$ for $x$ large enough.
We look for a similarity solution $F(x,t)=f(u)$, with a variable $u$ of the form
 $x/t^{\alpha}$. Replacement in eq.~(\ref{evolutionlin}) imposes that
\begin{equation}
\label{u}
u=3 \frac{x}{V^{\ast} \theta_{0}^{3}t} \ ,
\end{equation}
and that $f$ verifies
\begin{equation}
\label{f}
-f(u) + u f'(u) + 1=\frac{d}{du} \left(u^{3}f'''(u)\right).
\end{equation}
[The constants in eq.~(\ref{u}) have been chosen so as to cancel out all physical
quantities from eq.~(\ref{f})].

Since we require that for $t>0$, $e=0$ at $x=0$, we have the
boundary condition $f(u=0)=0$. Also, for the reasons exposed above, far away from the contact
line, we must have $f(u \rightarrow +\infty)= 1$. We finally
impose zero horizontal flow at $x=0$ and $x \rightarrow +\infty$.

The behavior of $f$ is shown on fig.~\ref{drop}a. For $u \rightarrow 0$, $f$
is not analytical. For $u \rightarrow +\infty$, $f$ presents decreasing oscillations, and behaves like a
linear combination of $e^{-\frac{3}{2}u^{1/3}} \cos \left(3 \frac{\sqrt{3}}{2} u^{1/3}\right)$
and $e^{-\frac{3}{2}u^{1/3}} \sin \left(3 \frac{\sqrt{3}}{2}
u^{1/3}\right)$. The essential feature of $f$ is that it
approximately reaches its asymptotic value ($f=1$) for $u$ of
order unity.

Therefore, we can conclude that the effect of the pinning is  felt only in a
region of breadth $\xi(t)=\frac{1}{3}V^{\ast} \theta_{0}^{3}t$, and
that beyond this ``healing length'', the drop recovers the profile
it would have adopted in the absence of contact line pinning.
Finally, the initial (macroscopic) contact angle of the drop is slightly
 reduced during the pumping:
\begin{equation}
\frac{\delta \theta}{\theta_{0}} \simeq \frac{J}{V^{\ast} \theta_{0}^{4}}
\end{equation}
[justifying in retrospect the linearization of eq.~(\ref{evolution})]. The global shape of
the drop is shown on fig.~\ref{drop}b.

\subsection{Pull out with hysteresis}

We now return to the modified Landau-Levich problem with the geometry of fig.~\ref{plate}b
and we allow for some hysteresis in the contact angle.

We start with the case of \textit{strong pinning}, where the receding
angle $\theta _{r}$ is equal to zero. Let us denote $F(V)$ the pinning force felt by the
contact line at velocity $V$. Strong pinning implies that, for a flat
macroscopic film, the force required to depin the contact line,
equal to  $F(V=0)$, is larger than the capillary force $\gamma _{SL}+\gamma$.
\footnote{ In our notation, the capillary force $\gamma_{SO}$ has been absorbed in $F(V)$.}
Formally, this is exactly similar to what we had in our discussion of section~\ref{wettable},
with a wettable surface displaying no hysteresis and $\gamma _{SO}>\gamma
_{SL}+\gamma$. The spreading parameter $S$ should simply be `renormalized'  to
$\tilde{S}=F(V)-\gamma_{SL}+\gamma$.
In the region close to the contact line, the liquid forms a thin
strip, with a thickness of the order of the pancake thickness $e_{p}$. Here,
$e_{p}$ is defined by the force balance
\begin{equation}
\label{newbalance}
F(V)-\left( \gamma _{SL}+\gamma \right) =P(e_{p})+e_{p}\Pi
(e_{p}).
\end{equation}
We do not know much about the velocity dependence of the pinning force $F(V
)$ at large $V$. For simplicity, we shall assume that $F(V)$ is an
increasing function of V. Thus $F(V)>F(0)$, and the left hand side of
eq.~(\ref{newbalance}) is constantly positive. Hence, eq.~(\ref{newbalance})
 admits, in principle, a solution $e_{p}(V)$.

Again $e_{p}$ is very small, and we expect the visible profile to be
dominated by the macroscopic solution: dissipation occurs only within a
finite vertical distance $\ell$. At greater distances, we return to a
uniform flow of tilt angle $\theta$.

Let us now turn to the case of \textit{weak pinning} ($\theta _{r}>0$).
Here, our pinning force $F(V)$ is weaker at low $V$:
\begin{equation}
F(0)-(\gamma _{SL}+\gamma )<0.
\end{equation}
But we are dealing with relatively high velocities $V$ at which $F(V)$ may
become larger. We are thus led to two possibilities:

\qquad a) if $F(V)>\gamma +\gamma _{SL}$, we return to the case of strong
pinning.

\qquad b) if $F(V)<\gamma +\gamma _{SL}$, we expect a profile similar to
what we discussed in section~\ref{partialwetting} for partial wetting:
at the approach of the contact line, the profile thickens, forming a
bump.

In both cases, again, the transition to the macroscopic flow occurs on
a microscopic region of size $\ell$.

\section{Concluding remarks}
\label{sec5}

We conclude with the following remarks:

1) A number of interesting experiments are conceivable with a pumped porous
medium: for instance, in pull out, measuring the finite height $H$ achieved
by a Landau-Levich film. The surprise is that $H$ is dictated by the uniform
tilted flow:

\begin{equation}
H \simeq \frac{e_{L}}{\theta } \simeq e_{L} \frac{V}{J} \; ,
\end{equation}
and is insensitive to the delicate dissipation processes taking place near
the triple line: the ``near'' field is of nanoscopic size for most cases.
And in practice, the finite size of the pores may provide a more important
cut off than the length $\ell$ discussed in sections~(\ref{wettable}) and
(\ref{partialwetting}).

2) There are interesting questions related to the pinning forces $F(V)$ at
large velocities. Here, the contact line is not necessarily the end
point of a liquid wedge: it may represent the border between a dry region
and a thin pancake. But the application of these ideas to porous media, with
pores of finite diameter, seems remote.

3) If we want to avoid these difficulties, we can think of pumping a liquid
droplet, on a non porous surface, by \textit{evaporation}: this was indeed
the starting point of a series of experiments by the Chicago group~\cite{gouttecafe}.
We benefit by having a smooth solid surface. But evaporation has its own
complications: temperature gradients induce Marangoni flows, which are
complex and not very instructive.

4) Returning to the pull out problem, we should also insist on another
limitation of our discussion: we restricted ourselves to steady state
regimes. At high velocities, the liquid wedge might decide to emit a
periodic train of droplets (or pancakes): these droplets would then be
pumped out by the current $J$. A tap delivering a low output of water is a
good example of periodic droplet emission. This idea is attractive. However,
because of the nanoscopic size of the objects involved, the effects may be
hard to observe.

\section*{Acknowledgements}
The authors thank F.\ Brochard-Wyart and J.-F.\ Joanny for fruitful discussions.


\newpage


\begin{figure}
\caption{Macroscopic puddle under aspiration.}
\label{puddle}

\end{figure}


\begin{figure}

\caption{Film energy versus film thickness in the case of:
  a) partial wetting,  b) complete wetting.}
\label{filmenergy}

\end{figure}


\begin{figure}

\caption{Pull out at velocity $V$ of a vertical plate from a bath:
a) The classical Landau-Levich problem,
b) the modified problem, with a porous plate inducing an aspiration current
$J$. Most of the film then displays a tilted profile of angle $\theta =
J/V$, with a plug flow velocity field, and terminates at a height $H=e_{L}/\theta$.
In the present article, we investigate
the microscopic structure of the film near the contact line (dotted circle).}
\label{plate}

\end{figure}


\begin{figure}

\caption{Film profile in the complete wetting case, for $\epsilon=2.3 \; 10^{-5}$ and
$S/\gamma = 0.2$ (length unit is $\ell$ on both axes). (a) Vicinity of the
contact line. (b) Outer region. After a few $\ell$'s, the film merges with the macroscopic
tilted flow of angle $\theta$ (asymptote). Note, in each plot, the very different
scales between the two axis : the one giving the distance from the contact
line should be considerably stretched to get a representative idea of the
actual film profile.}

\label{wettable}

\end{figure}


\begin{figure}

\caption{Film profile near the contact line in the partial wetting case,
for $\theta = 10^{-3}$ and $\epsilon = 2.3 \; 10^{-5}$ (length unit is $\ell$ on both axes).
The profile presents a `bump' corresponding to the intermediate region. Farther away
from the contact line, the profile approaches the asymptote in a way similar
to the wetting case (not shown).}
\label{partialwetting}

\end{figure}


\begin{figure}

\caption{Bump steepening in the film profile (partial wetting) as the pull-out
velocity is increased (decreasing values of $\epsilon$).}
\label{bumpsize}

\end{figure}


\begin{figure}

\caption{Macroscopic, pinned, drop under aspiration. (a) Shape of the function
$f(u)$ (see text). (b) Deformation of the liquid wedge near the line. The grey
line represents the initial profile. The effect of pinning is felt inside a
region of width $\xi(t)$; beyond, the profile (solid line) ``heals'' and recovers
the shape it would have had in the absence of pinning (dotted line).}
\label{drop}

\end{figure}

\end{document}